\documentclass[preprint2,twoside]{hwo}

\usepackage{lipsum}

\input{hwo.h}

\begin{document}

\title{\textbf{\LARGE WaveDriver: a Laser Guide Star AO System for HWO}}
\author {\textbf{\large Benjamin L.\ Gerard,$^1$ Alex Geringer-Sameth,$^1$ Aditya R.\ Sengupta,$^2$ Alexx Perloff,$^1$ Michael Messerley,$^1$ Dominic F.\ Sanchez$^1$, P.\ Waswa$^1$, William Moore,$^1$ Matthew Cook,$^1$ Eric Strang,$^1$ Paul Pax,$^1$ Cesar Laguna,$^1$ Matthew DeMartino,$^2$ Kevin Bundy,$^2$ Rebecca Jensen-Clem,$^2$ Aaron J.\ Lemmer,$^1$ S.\ Mark Ammons,$^1$ Lisa Poyneer,$^1$ Megan Eckart$^1$}}
\affil{$^1$\small\it Lawrence Livermore National Laboratory, CA, USA}
\affil{$^2$\small\it University of California Santa Cruz, CA, USA}



\begin{abstract}
HWO’s Tier 1 Contrast Stability Technology Gap presents a key challenge for technology development in the coming years, requiring to a $>$ $100\times$ more stable system than \textit{JWST}. WaveDriver is a concept for a laser guide star spacecraft coupled to an adaptive optics (AO) system onboard HWO that would enable HWO to reach its picometer-level wavefront stability requirements while relaxing other HWO subsystem requirements. At LLNL and UCSC we are revisiting the concept initially proposed by Douglas et al.\ (2019). We present results from our project’s first year, including (1) AO control developments, including with Linear Quadratic Gaussian control and machine learning, (2) AO wavefront sensor (WFS) trade study simulations, and (3) simulations, fabrication, and testing of a 133-port photonic lantern WFS/spectrograph. A key finding from our work is that WaveDriver could be needed to enable HWO's primary mirror segment stability and/or low order wavefront stability requirements.
  \\
  \\
\end{abstract}

\vspace{2cm}
\vspace{-10pt}
\section{Introduction}
\begin{figure}[h]
    \centering
    \vspace*{-5pt}
    \includegraphics[width=1.0\linewidth]{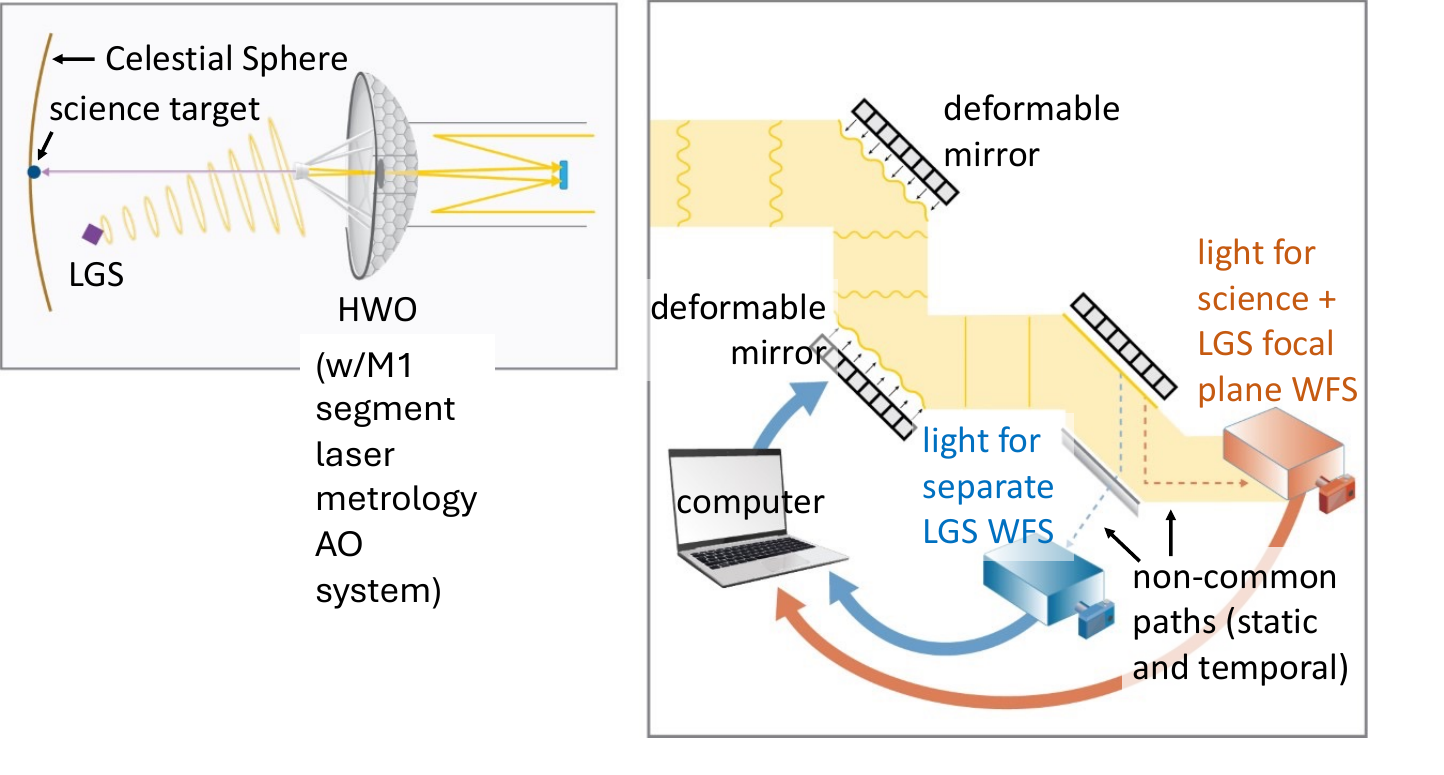}
    \vspace*{-20pt}
    \caption{Conceptual Illustration of the HWO Laser Guide Star (LGS) adaptive optics (AO) system concept considered here, called WaveDriver, designed to enable meeting HWO's wavefront stability requirement of 10 pm rms over 10 minutes per mode, hereafter $\sigma_{10}$.}
    \vspace{-8pt}
    \label{fig:concept}
\end{figure}

\vspace{-5pt}
\label{sec:intro}
Work that led to \cite{Douglas2019}, \cite{Clark2020}, and \cite{Poter2022} has considered the concept of a laser guide star (LGS) spacecraft to illuminate a Habitable Worlds Observatory (HWO)-like telescope to enable adaptive optics (AO) correction for enhanced telescope wavefront stability. We revisit and build on these developments here explicitly for HWO, now called ``WaveDriver'' (naming credit: Keith Jahoda) and illustrated in Fig.~\ref{fig:concept}.
Note that the prior work referenced above does not explicitly consider the joint use of HWO's existing internal laser metrology system for primary mirror segment stabilization \citep{Redding2024}, whereas here, as shown in Fig.~\ref{fig:concept}, we consider this in combination with WaveDriver. Here we consider using a photonic lantern (PL) as the natural guide star (NGS) wavefront sensor (WFS) in \S\ref{sec:pl}, LGS WFS trade study simulations in \S\ref{sec:wfs}, and AO Control simulations in \S\ref{sec:control}.
\vspace{-10pt}
\section{PL as a NGS WFS}
\begin{figure}[h]
    \centering
    \includegraphics[width=1.0\linewidth]{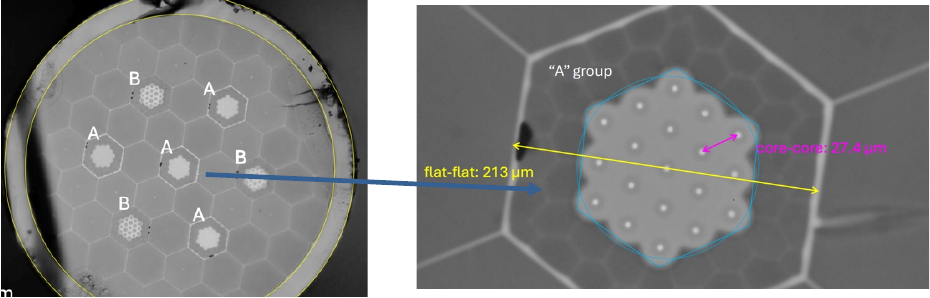}
    \vspace*{-20pt}
    \caption{Microscope images of the single mode end a photonic lantern recently fabricated at LLNL, which includes a bundle of seven multi-mode inputs (decreased relative to the dimensions in this figure by a taper ratio of 7.14) each of which produces 19 single-mode outputs.}
    \label{fig:pl}
    \vspace{-5pt}
\end{figure}

\vspace{-5pt}
\label{sec:pl}
In the context of Fig.~\ref{fig:concept}, we consider a photonic lantern (PL; \citealt{Birks2015}) as a NGS WFS in WaveDriver, motivated by the need for, at minimum, NGS tip/tilt/focus measurements that would be discrepant with equivalent LGS WFS measurements due to WaveDriver spacecraft motion. We have tentatively determined that the HWO photon noise-limited WFS measurement error reaches $\sigma_{10}$ for tip/tilt/focus with a four port PL at $\text{m}_{\text{V}}\sim3.6$, which will be described in detail in a future paper. We have recently fabricated 133-port PL at LLNL, recently initially presented in \cite{Sengupta2025} and also shown here in Fig.~\ref{fig:pl}.
The PL was fabricated in a multi-step process using two different single-mode pre-forms, one producing the more classical design ``A'' from OFS and the other ``B'' pattern producing a so-called depressed-well design from Weatherford (Preform 190916-1). Designs A and B are denoted accordingly in Fig.~\ref{fig:pl}. The B design preform uses two different fiber core materials and with three different core sizes throughout the 19 cores (6 large, 6 medium, 7 small). Both bundles are designed for a 633-nm central wavelength, but the B design provides a potentially larger spectral range over which single mode behavior exists but at the expense of decreased throughput. More details on the design and fabrication of this PL will be described in a future paper. Efforts to characterize this PL with off-axis holography and via a 133-port lenslet-based integral field spectrograph are ongoing, being led by co-authors Sengupta and Sanchez, respectively, the results of which will be published in a future paper. 
\section{LGS WFS Simulations}
\label{sec:wfs}
\begin{figure}[!h]
    \centering
    \includegraphics[width=1.0\linewidth]{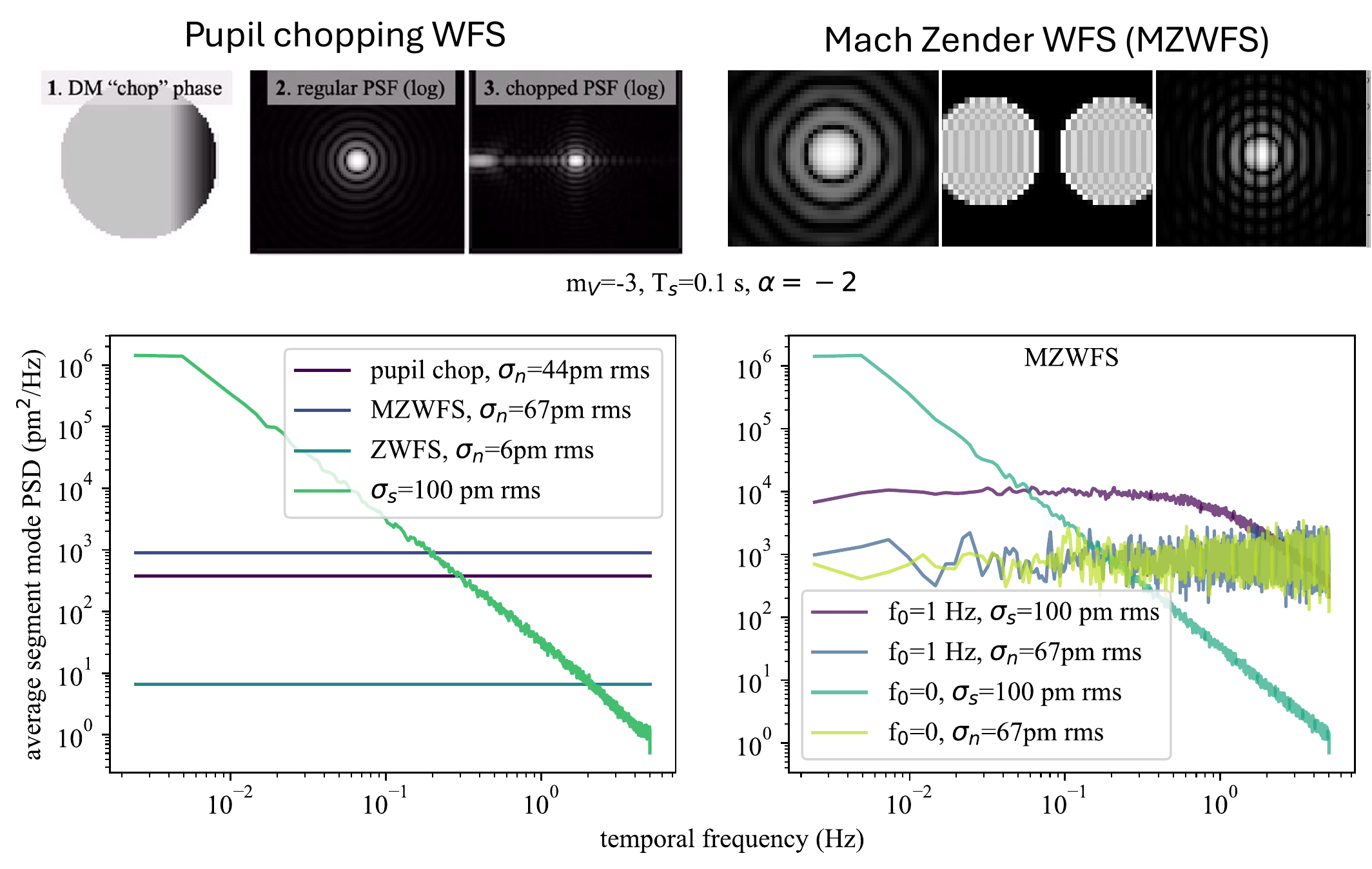}
    \caption{LGS WFS trade study results, showing (1) a Zernike WFS (ZWFS) is the most sensitive WFS considered here, (2) a LGS ($\text{m}_{\text{V}}<-3$) is needed to reach HWO's $\sigma_{10}$ requirements for M1 segment stability, and (3) the photon noise limits ($\sigma_n$) determined here are not dependent on any input power spectral density shape assumptions.}
    \label{fig:lgs_wfs}
\end{figure}

\cite{Douglas2019} considered the use of WaveDriver with a Zernike WFS (ZWFS) for primary mirror (M1) segment stabilization via active M1 segment wavefront control. We revisit this topic here by comparing additional WFSs to the ZWFS and also considering the need for correction of lower spatial frequency errors with a separate deformable mirror (DM). We used the HWO exploratory analysis case 1 (EAC1) M1 configuration with 37 unobscured hexagonal segments and considered piston+tip+tilt modes for each segment and parameters otherwise matched to \cite{Douglas2019}, which were then Fraunhoffer-propagated to form various WFS images. The DM-based pupil chopping WFS \citep{Gerard2023, Soto2023} and a Mach-Zhender WFS were simulated in addition to a ZWFS, shown in Fig.~\ref{fig:lgs_wfs}.
The simulations used to generate this figure inject M1 segment temporal disturbances with a given power law, $\alpha$, and spectral break, $f_0$, and an input modal wavefront error (WFE), $\sigma_s$, of 100 pm rms, following the ``worst case'' scenario in integrated modeling results from \cite{Poter2022}. The time domain wavefront data cubes are generated at a cadence of 0.1s and run for 10 minutes in simulations both with and without photon noise to produce a 6000$\times$111$\times$2 array of M1 segment modal coefficients. The difference of coefficients with vs.\ without photon noise, then averaged over all 111 M1 segment modes, is what produces the various $\sigma_n$ curves in Fig.~\ref{fig:lgs_wfs}, for which the corresponding number in the legend is computed by taking the square root of the integral of each plotted curve. These open loop photon noise limits can be thought of as limits to achievable closed loop wavefront error with optimal linear control (see \S\ref{sec:control}) and can thus be considered fundamental limits for the purposes of this analysis.

This same simulation framework is used to simulate photon noise limits for injected low order error wavefront errors, for which we find with a ``perfectly achromatic'' Zernike WFS (i.e., flux scaling with V band but with no PSF magnification and/or scalar mask chromaticity, representing an upper limit on achievable WFS sensitivity) on a $\text{m}_{\text{V}}=3$ star (i.e., among the brightest targets for HWO), that with $\sigma_s=100$ pm rms per mode $\sigma_n=18$ pm rms for an average the first 15 Zernike modes. This suggests that a LGS is likely also needed to reach $\sigma_{10}$ for low order modes. 
\section{Control Simulations}
\label{sec:control}
\begin{figure}[h]
    \centering
    \includegraphics[width=1.0\linewidth]{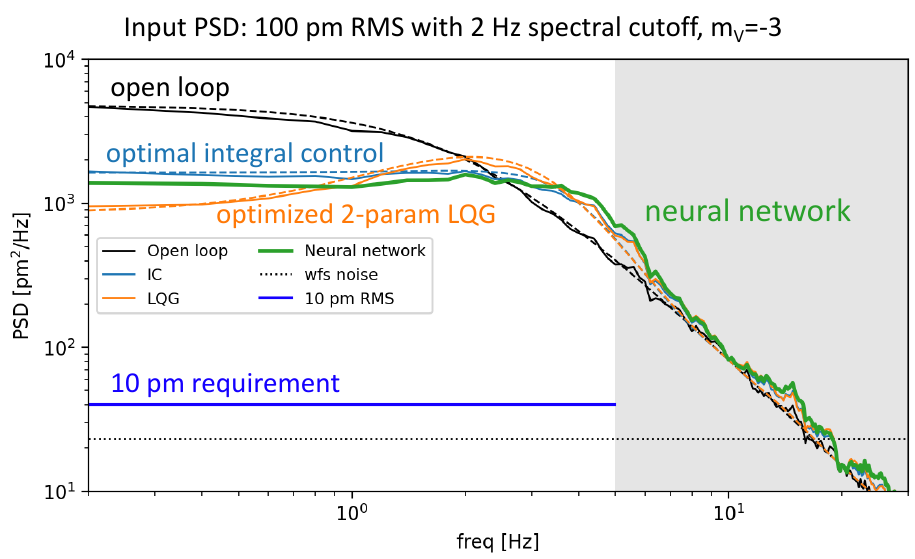}
    \vspace*{-25pt}
    \caption{AO bandwidth error may prevent the HWO M1 laser metrology wavefront control system, which has a 10~Hz actuation limit \citep{Redding2024}, from reaching HWO's 10 pm requirement ($\sigma_{10}$). The input ``open loop'' PSD reflects integrated modeling results from \cite{Poter2022}, with various closed-loop PSDs (light blue, green, orange) clearly not reaching $\sigma_{10}$ (dark blue), despite the measurement noise (dotted black) being below $\sigma_{10}$.}
    \label{fig:control}
\end{figure}

\begin{figure}[h]
    \centering
    \includegraphics[width=1.0\linewidth]{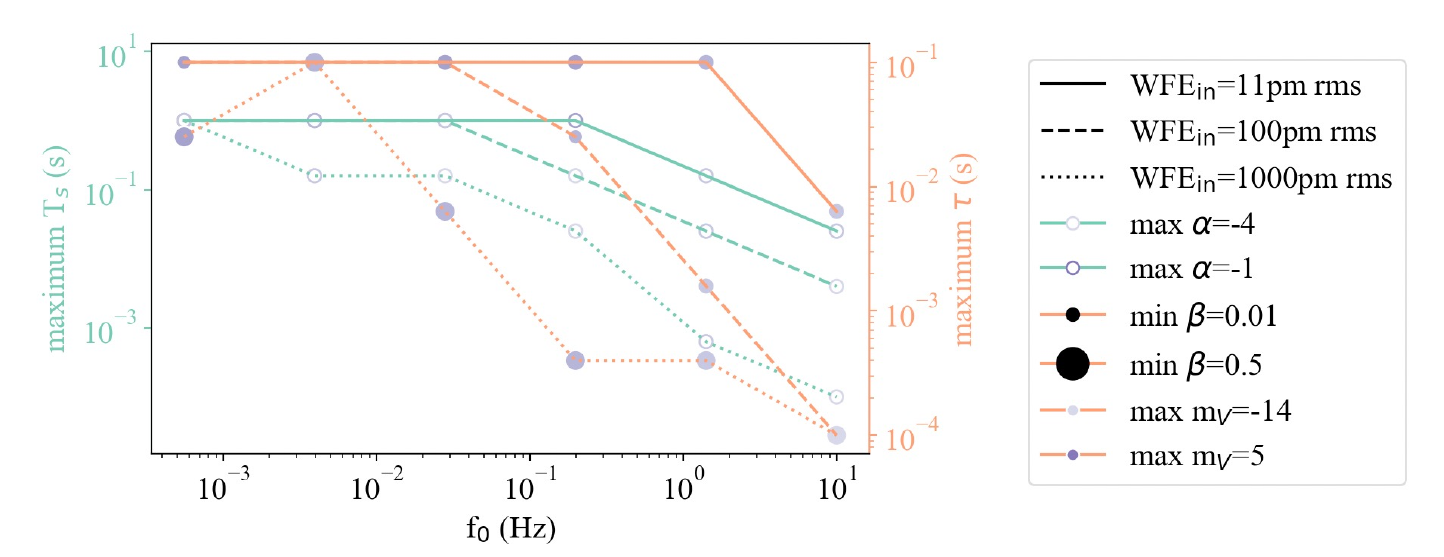}
    \vspace*{-20pt}
    \caption{The required AO system parameters---frame rate ($T_s$), latency ($\tau$), and WFS sensitivity ($\beta$; \citealt{Guyon2005})---needed to reach $\sigma_{10}$ in response to input disturbance parameters---integrated rms stability between 1/(10 minutes) and the WFS' temporal Nyquist limit (WFE$_\mathrm{in}$), spectral break ($f_0$), and high frequency power law ($\alpha$)---and guide star magnitude ($m_V$).}
    \label{fig:limits}
\end{figure}

We conduct closed-loop AO simulations to consider the combined effects of temporal bandwidth error and measurement noise in limiting or not the reaching of $\sigma_{10}$ (with \S\ref{sec:wfs} thus far only considering measurement noise). As in \S\ref{sec:wfs} we draw from \cite{Poter2022} to consider an input power spectral density (PSD) with $\sigma_s=100$ pm rms with a $f_0=$2 Hz spectral cutoff and a spectral slope of $\alpha=$-2, using variables consistent with equation 2 in \cite{Douglas2019}. To address potential AO bandwidth errors, we developed an AI-based controller using reinforcement learning (RL) and assuming the use of an FPGA. We used FPGA-benchmarking software (to be described in more detail in a future paper) to determine that realistic computational latency for M1 segment wavefront control calculations using a circular buffer with the past 100 frames as an input into the RL controller at each new WFS frame would be of order a few $\mu$s, thus using a total AO system latency based on typical performance (including WFS camera read out, WFS camera frame transfer, computation, DM frame transfer, and DM actuation) of 0.5 ms for which bandwidth error is dominated by the 0.1-s exposure time limit. This latency and and 10 Hz frame rate were tuned on a perfect predictive controller and are compared to an optimal leaky integral controller \cite{Gendron94} and optimal Linear Quadratic Gaussian auto-regressive order 2 controller using the framework from \cite{Poyneer2023}, all of which are shown in Fig.~\ref{fig:control}.
The shaded region in Fig.~\ref{fig:control} is above the 5 Hz Nyquist limit of the M1 laser metrology wavefront control system \citep{Redding2024}, but a higher temporal bandwidth high order DM and/or segmented DM (i.e., illuminated by WaveDriver) could potentially reduce this bandwidth error below $\sigma_{10}$. To assess this potential solution, we developed and open-sourced the following code: \url{https://github.com/LLNL/hwo1dGUI}. This code was ultimately used to produce Fig.~\ref{fig:limits}.
Exploring parameters in the GUI and Fig.~\ref{fig:limits} both illustrate the following key findings of when/how $\sigma_{10}$ can be achieved:
\begin{enumerate}
    \item $\text{m}_{\text{V}}\lesssim -3$, corroborating the conclusions from \cite{Douglas2019}, \cite{Poter2022}, and \cite{Redding2024} that a bright laser source is needed to stabilize the segments, with science target stars being fundamentally too faint to reach $\sigma_{10}$, leaving WaveDriver (external) and laser metrology (internal) as possible LGS solutions.
    \item when the input WFE is ``passively'' stable below 10~pm rms. This can be equivalently re-stated as requiring passive stability of 8 pm rms between 1 and 5 Hz, where this temporal frequency range is above the accessible M1 segment laser metrology temporal bandwidth, thus requiring passive stability at these high frequencies not accessible to attenuation via active control. Recent HWO EAC1 integrated modeling presented at this conference where micro-thrusters were used instead of gyroscopes \citep{Feinberg2025, Zeimer2025} suggests that this is achievable and is the baseline strategy for reaching $\sigma_{10}$ (Feinberg, Liu, et al., private communication).
    \item when integrated input disturbances are above 10 pm rms but the spectral break ($f_0$) and/or power law ($\alpha$) are adjusted to shift power to lower temporal frequencies that are below the $\sim$1 Hz M1 segment bandwidth limit. For example, for $\sigma_s\equiv\text{WFE}_\mathrm{in}=100$ pm rms, $\alpha=-2$, and $f_0<0.25$ Hz, $\sigma_{10}$ can be reached with $T_s=0.1$ s, $\tau=1.5$ ms. This indicates an area potentially worth investigating where system-level telescope design optimization can potentially shift input high frequency input disturbances to lower frequency output disturbances.
    \item\label{pt:wavedriver} using a ``second stage'' high temporal bandwidth DM, which would require WaveDriver. For the 100 pm rms input case from \citealt{Poter2022} (additionally with $f_0=1$ Hz and $\alpha=-2$), $\sigma_{10}$ can be achieved with $\tau=1.5$ ms for $T_s \lesssim 20$ ms, which a DM could enable. Although HWO-relevant DM actuation speeds have thus far been limited to speeds $\lesssim$10~Hz to meet voltage stability requirements (e.g., \citealt{Bendek2020, Bendek2024}), there is agreement from ongoing discussion with DM electronics subject matter experts that development to enable $\gtrsim$100 Hz actuation speeds while still meeting voltage stability requirements is potentially feasible and worth investigating (G.\ Ruanne, E.\ Bendek, T.\ Groff, private communication). Note that in principle the inherent closed-loop nature of a WaveDriver AO system would relax DM votage stability requirements at frequencies below the expected WaveDriver temporal bandwidth ($f\lesssim9$ Hz for the example given at the beginning of this bullet point). Also note that it has sometimes been stated that laser metrology ``feed-forward control'' at $>$ 100 Hz frame rates could be enabled with a high bandwidth continuous and/or segmented DM, but this nomenclature is misleading as in reality this would be inherently open loop control, which would (1) add risks of calibration errors and (2) prevent the typical 2 dB/decade of rejection at low temporal frequency (i.e., a controller with gain = 1 and leak = 0 does not provide such rejection) and thus add additional constraints on other parameters such as maximum input WFE, latency, frame rate, etc.
\end{enumerate}
\section{Conclusion and Discussion}
\label{sec:conclusion}
Wavefront stability such that contrast is maintained at 10$^{-10}$ levels over a science exposure is a key technical requirement for HWO and potentially not currently feasible with current state-of-the art technologies. In this proceedings we build off of previous work (\citealt{Douglas2019}, \citealt{Poter2022}) to propose WaveDriver, a LGS AO system for HWO, as a solution to bridge this contrast stability technology development gap. We fabricated a 133-port photonic lantern, presented in \S\ref{sec:pl}, as a proposed NGS WFS in the WaveDriver system. We considered different WaveDriver LGS WFS options in \S\ref{sec:wfs}, confirming that a Zernike WFS is the most sensitive to photon noise and that a LGS is inherently needed for segment stabilization at $\sigma_{10}$ levels. Lastly in \S\ref{sec:control} we investigated potential machine learning and FPGA-based AO control options for WaveDriver and in doing so found a potential $\sigma_{10}$ temporal bandwidth error limitation from M1 segment and/or low order wavefront errors, for which a potential solution requiring WaveDriver that would use a high bandwidth DM was proposed. 
Expanding further on point \ref{pt:wavedriver} in \S\ref{sec:control}, we are aware that the baseline HWO strategy for high temporal frequency M1 segment stabilization at will be to engineer the telescope to meet $\sigma_{10}$ requirements ``passively,'' but it is worth noting here that even if WaveDriver is considered a backup plan that a risk assessment should ultimately help guide this possible WaveDriver application; e.g., using the standard systems engineering risk matrix, if there is a $>$10\% chance of WFE$_\mathrm{in}>8$ pm rms between 1 and 5 Hz, not having WaveDriver as a solution to address this would in principle result in at least medium risk to HWO, which should be evaluated against HWO's overall risk tolerance. Also note that point \ref{pt:wavedriver} applies to both high and low order errors, and thus use of WaveDriver for high-speed active correction of low order errors is still of interest for future HWO technology development (T.\ Groff, private communication). We are also aware that WaveDriver is of interest for low temporal frequency absolute phase correction of M1 segment phasing error to increase sky-coverage relative to the out-of-field NGS sensing (A.\ Jurling, private communication).
\acknowledgements
{\bf Acknowledgements.} We thank Michael (Mike) Messerley, who passed away unexpectedly this past spring, for his interest, enthusiasm, and innovation that led to the fabricated photonic lantern presented in this paper. Mike, your work has inspired the current and next generation of scientists and engineers. We also thank Keith Jahoda, Aki Roberge, Ewan Douglas, Kerri Cahoy, Jared Males, and Olivier Guyon for discussions that led to the development of this project. We also thank Garreth Ruanne, Jonathan Tesch, Breann Sitarski, Michael McElwain, Mitchell Troy, Alice Liu, Alden Jurling, and Tyler Groff for discussions that informed the content in this proceeding. This document number is LLNL-PROC-2011254. This work was performed under the auspices of the U.S.\ Department of Energy by Lawrence Livermore National Laboratory under Contract DE-AC52-07NA27344.  

\bibliography{author.bib}

\end{document}